\documentclass[10pt,letterpaper]{article}
\usepackage{opex3}
\usepackage{color}
\usepackage{cite}
\usepackage{ifpdf}
\begin{document}

\title{Autonomous absolute calibration of an ICCD camera in single-photon detection regime}

\author{Luo Qi,$^{1,2,*}$ Felix Just,$^{1,2}$ Gerd Leuchs,$^{1,2}$ and Maria V. Chekhova$^{1,2,3}$}

\address{$^1$Max-Planck Institute for the Science of Light, G.-Scharowsky Str 1/Bldg 24, 91058 Erlangen, Germany\\
$^2$Friedrich-Alexander University of Erlangen-N{\"u}rnberg, Staudtstrasse 7/B2, 91058 Erlangen, Germany\\
$^3$5Department of Physics, M. V. Lomonosov Moscow State University, Leninskie Gory, 119991 Moscow, Russia}

\email{$^*$luo.qi@mpl.mpg.de} 



\begin{abstract*}
Intensified charge coupled device (ICCD) cameras are widely used in various applications such as microscopy, astronomy, spectroscopy. Often they are used as single-photon detectors, with thresholding being an essential part of the readout. In this paper, we measure the quantum efficiency of an ICCD camera in the single-photon detection mode using the Klyshko absolute calibration technique. The quantum efficiency is obtained as a function of the threshold value and of the wavelength of the detected light. In addition, we study the homogeneity of the photon sensitivity over the camera chip area. The experiment is performed in the autonomous regime, without using any additional detectors. We therefore demonstrate the self-calibration of an ICCD camera.
\end{abstract*}



\section{Introduction}

An intensified charge coupled device (ICCD) camera is one of the promising types of matrix detectors with single-photon sensitivity. In general, an ICCD camera consists of a photocathode, a micro-channel plate (MCP), a phosphor screen, a CCD imaging chip accompanied by readout electronics, and an optical coupling system between the latter two. Such a design allows vast flexibilities: for instance, by selecting photocathode materials, one can achieve higher sensitivity in a preferred spectral range~\cite{siegmund1999microchannel,mende2000far}. The high voltage applied to the MCP performs not only as an electronic multiplier but also like an electronic shutter, and the temporal dynamics can be as fast as on nanosecond scale. Moreover, by tuning the voltage amplitude of MCP, the signal amplification factor (gain) can be varied to adapt for the light sources of different brightness. Since the late 1970s, when the first ICCD cameras were developed, the family of ICCD cameras has been widely implemented in astronomy and space research~\cite{fordham1989astronomical,mende2000far,torr1986intensified,timothy2013microchannel}, chemistry, biology, and medicine~\cite{rettig2012applied}. Nowadays, ICCD cameras are more and more often used in quantum optics~\cite{haderka2005direct,fickler2013coincidence,fickler2013real,aspden2013epr,tasca2013influence,morris2015imaging,shcherbina2014photon}. Especially, because an incident light beam can be spread over a large number of pixels, an ICCD camera can be considered as a multi-channel photon-number resolving detector, albeit with low quantum efficiency. There are several other types of cameras, which have similar applications, for instance EMCCD~\cite{blanchet2008measurement,edgar2012imaging,brida2010experimental,avella2016absolute} and intensified sCMOS cameras (I-sCMOS)~\cite{chrapkiewicz2014high,jachura2015shot}. Compared to an EMCCD camera, an ICCD camera has significantly lower dark noise~\cite{buchin2011low,schuhle2013intensified}, which is essentially resulting in the technique of counting single photons one by one, not possible for an EMCCD camera. As to I-sCMOS cameras, they have the same design as the ICCD cameras but with the CCD chip replaced by a sCMOS chip.

An essential part of detecting single photons with an ICCD camera is distinguishing a single-photon event from the dark noise. As in most photon counting devices, this is done by applying a threshold for the charge read out from the output. Typically, this charge has an exponential probability distribution for the cases of both dark noise and the incident photons at the input. For this reason, the value of the threshold affects both the signal-to-noise ratio (SNR) and the quantum efficiency (QE). For threshold values allowing a sufficiently high signal-to-noise ratio, the quantum efficiency is therefore significantly less than expected from the photocathode datasheets.

Here we measure the quantum efficiency of an ICCD camera operating in the single-photon detection regime, as a function of the threshold value. The measurement is performed using the absolute calibration method~\cite{klyshko1980use,malygin1981absolute,ware2004single,just2014detection}, with the radiation of spontaneous parametric down-conversion (SPDC) at the input. The absolute calibration technique is applied in an autonomous manner, i.e., with the ICCD camera used as both the reference detector and the detector under test (DUT). We also study the quantum efficiency versus the wavelength of the incident radiation as well as its uniformity over the camera matrix.

\section{Single-photon detection with ICCD}

\begin{figure}[htb]
\centering
\includegraphics[width=10cm]{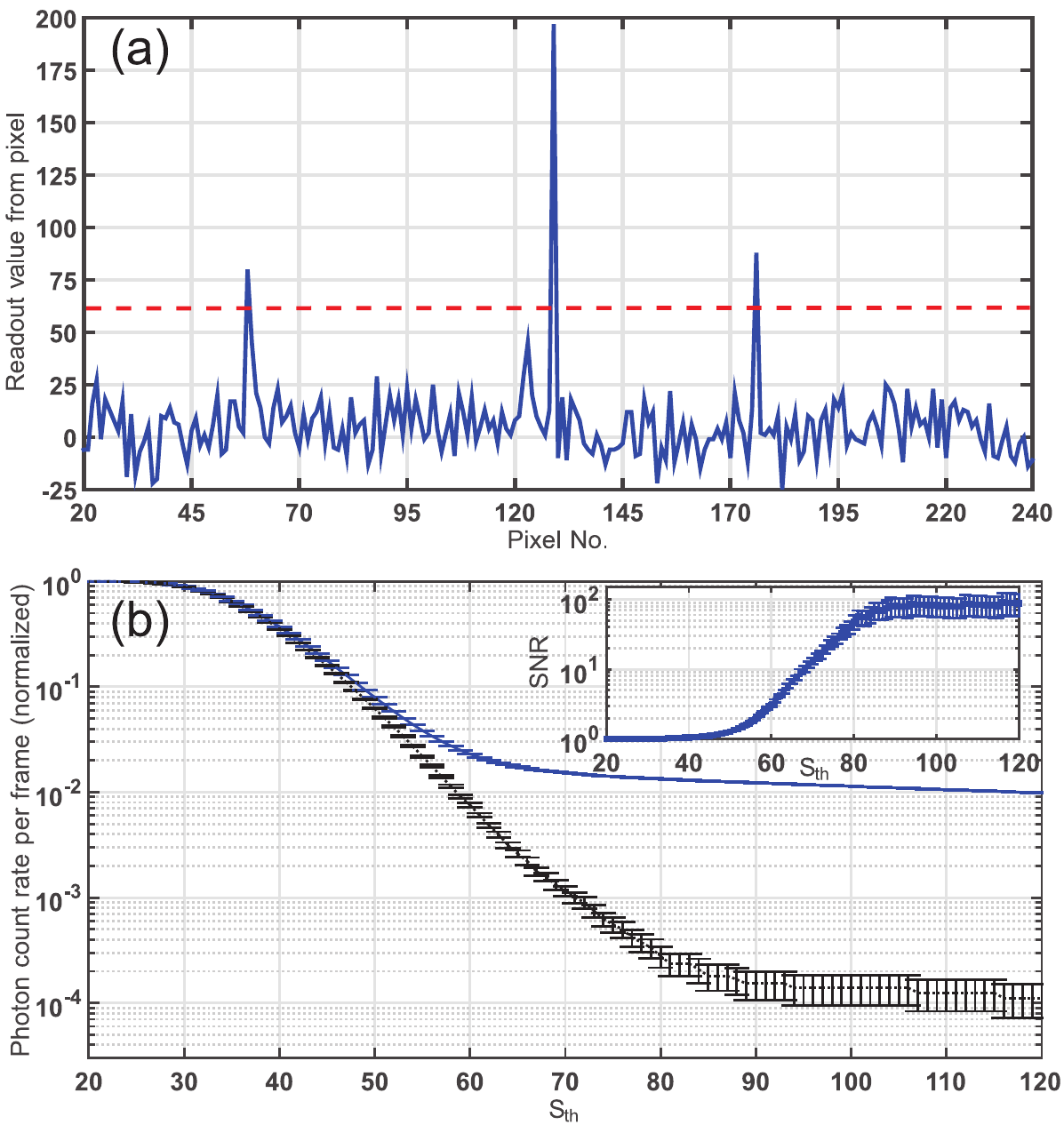}
\caption{(a) The readout values of an array of pixels in the ICCD camera (blue solid curve) and the threshold parameter $S_{th}$ (red dashed line). Every peak that is above the threshold is interpreted as a photon detection event. (b) The mean number of photon detections on a single pixel per $100 ns$ gates with SPDC radiation (blue solid line) and no radiation (black dashed line) at the input as functions of the threshold value. Inset: Signal-to-noise ratio (SNR) as a function of the threshold value.}\label{Fig1}
\end{figure}

In the single-photon detection mode, the ICCD camera works as a gated on-off detector. During the gate time, high voltage is applied to the MCP, and once an incident photon creates an electron on the photocathode, this electron will be multiplied in the MCP and create a light flash on the phosphor screen, which, in turn, creates a charge in some pixels of the CCD chip, although due to the inevitable crosstalk, it is never a single pixel that is affected by a single photon at the input. The CCD charge is finally read out in the digitized way. This value is somewhat higher than the background values, which mainly come from the electronic thermal noise on the CCD and the background luminescence of the phosphor screen. As illustrated in Fig.~\ref{Fig1}a, after reading out an array of pixels (blue solid curve), several pixels show higher values (peaks) than the background (fluctuations between $-25$ and $25$ for the readout value, after subtracting the average background level). In this situation, one uses a variable threshold parameter $S_{th}$ to distinguish the signal caused by the incident photon from the background noise on every pixel of the camera (Fig.~\ref{Fig1}a). It should be emphasized that this strategy is a binary processing and cannot accurately count how many photons are detected by the same pixel within one exposure time. Hence it is necessary to reduce the input photon rate until the existence of double-click event on a single pixel becomes negligible. Technically, this can be done by reducing the exposure time. Importantly, the thresholding process takes place after the data acquisition, and hence the threshold value can be chosen after the actual experiment.

In order to obtain single-photon detection events on an ICCD camera, it is important to choose the threshold correctly. For example, Fig.~\ref{Fig1}b shows two statistical measurement results (number of counts per exposure) of the signal (SPDC) and of the dark noise as functions of the threshold value. Note that here and further on, the threshold values are shown after the subtraction of the `background' value, which was obtained in the absence of illumination. This value varied from pixel to pixel within the range $600$ to $650$. One can see that, as the threshold value increases, the noise reduces faster than the signal, hence the signal-to-noise ratio (SNR) grows, as shown in the inset in Fig.~\ref{Fig1}b. But above the threshold of roughly $80$, SNR grows very slowly and practically reaches a constant value. This behavior can mean that the dark noise is almost completely suppressed, and the remaining noise results only from the stray light of the environment. In this threshold range, the detection events are reliable.

\section{Autonomous absolute calibration of the camera}

In addition to reducing the noise, an increase in the threshold value leads to the reduction of the quantum efficiency (QE). For this reason we performed an absolute calibration experiment on an ICCD camera using the Klyshko method~\cite{klyshko1980use,malygin1981absolute,ware2004single,just2014detection}, which is based on twin-photon correlations in SPDC. The method is absolute since it requires no knowledge of the reference detector QE for calibrating an unknown detector, so-called detector under test (DUT). In the case of pulsed radiation, the QE of the DUT can be expressed as~\cite{kwiat1994absolute}

\begin{equation}\label{Eq1}
\eta_{DUT}=\frac{N_{cc}-N_{acc}}{N_{Ref}-\Delta{N}_{n}}\\
\end{equation}
where $N_{Ref}$ is the number of photons per pulse registered by the reference detector, $\Delta{N}_n$ is the mean number of noise counts registered by the same detector per pulse, $N_{cc}$ is the mean number of coincidence counts per pulse between the reference and the DUT detector channels, and ${N_{acc}}$ is the mean number of the accidental coincidence counts per pulse. Based on this equation, we have measured the QE of an ICCD camera as a function of the threshold value and the wavelength of the detected light. In addition, we have tested the uniformity of the QE over the camera by measuring different pixels.

\begin{figure}[htb]
\centering
\includegraphics[width=13cm]{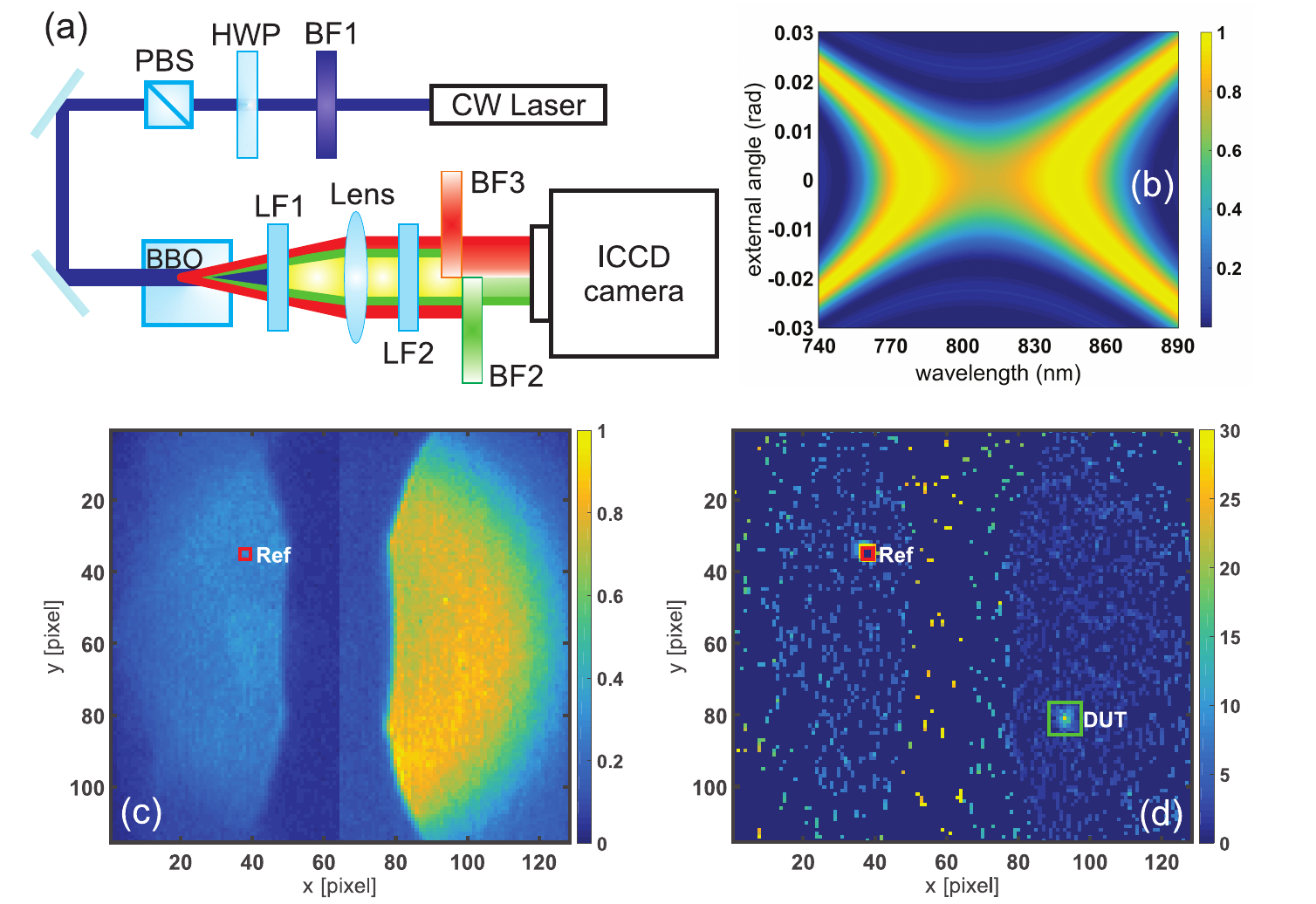}
\caption{(a) Experimental setup: a CW diode laser beam is filtered by a bandpass filter BF1 and used to pump SPDC in a $3$ mm BBO crystal; combination of a half-wave plate (HWP) and a polarizing beam-splitter (PBS) is used to control the pump power and to measure the noise caused by fluorescence and stray light. A long-pass filter LF1 suppresses the pump beam after the crystal; the SPDC radiation is far-field imaged onto the ICCD camera with a lens. After the lens, a long-pass filter LF2 filters out irrelevant wavelengths, and two bandpass filters (BF2 and BF3) cover different halves of the camera field of view for choosing the `reference' and the `DUT' wavelength bands. (b) Wavelength-angular spectrum for the collinear slightly non-degenerate SPDC. (c) A typical image taken by ICCD for the gate time $500~{\mu}$s, with the `reference' filter BF2 selecting a bandwidth of $10$ nm around $780$ nm and the `DUT' filter BF3, a bandwidth of $40$ nm around $850$ nm. The pixels inside the red square are chosen as the `reference' detector. (d) The result of the $g^{(2)}$ measurement between the `reference' detector and all the other pixels of the camera, the threshold being $80$. The green square shows the chosen `DUT' detector.}\label{Fig2}
\end{figure}

The experimental setup (Fig.~\ref{Fig2}a) uses type-I SPDC with collinear, close to frequency degenerate, phase-matching (see Fig.~\ref{Fig2}b) in a BBO crystal pumped by a CW diode laser beam (wavelength $405$ nm, mean power $80$ mW). Because of its broadband background spectrum, the pump is additionally filtered by a bandpass filter (BF1, CWL $405$ nm, FWHM $10$ nm, and transmission $95\%$). A long-pass filter (LF1, cut-off wavelength $450$ nm, transmission $>95\%$) after the BBO crystal reflects back the pump beam. The SPDC radiation is far-field projected onto an ICCD camera (PI-MAX3, Princeton Instruments) with the help of a lens (transmission $>99\%$) with $250$ mm focal length. After the lens, another long-pass filter (LF2, cut-off wavelength $650$ nm, transmission $>95\%$) is used to further suppress the SPDC radiation of irrelevant wavelengths.

For achieving the highest photon sensitivity, the MCP gain of ICCD camera is set to $100$. The ICCD camera is gated by switching on the MCP high voltage. Therefore, even though we use a CW pump, the photon count rate per frame, which is measured by the chosen detectors (groups of pixels) is fully equivalent to the photon count rate per pulse, which is required in Eq. (\ref{Eq1}). Similarly, the gate time is equivalent to the coincidence window~\cite{pittman1995optical,bennink2002two}. As mentioned above, the ICCD camera can count single-photon events correctly only under the condition of low input photon rate. According to this, we choose the gate time to be $100$ ns. We acquire more than 3 million frames for each experiment in order to obtain sufficient photon statistics. The binning is chosen to be $8\times8$ pixels, which leads to the acquisition rate as high as $\sim26$ Hz.

A combination of a half-wave plate (HWP) and a polarizing beam-splitter (PBS) is used to control the pump power and to measure the noise caused by fluorescence and stray light. When we rotate the pump polarization by $90\,^{\circ}$, the phase-matching condition is no longer satisfied and hence no SPDC light can be generated, but the level of fluorescence and stray light is the same. Therefore, the number of detected events for the $90\,^{\circ}$ orientation of the HWP is used as $\Delta{N}_n$ in Eq. (\ref{Eq1}). It includes all sources of noise: dark noise, fluorescence, and ambient light.

In order to calibrate the ICCD camera in an autonomous regime, we use its different groups of pixels as the reference detector and DUT. In the setup, we place two different filters in front of the input window of the ICCD camera, one (BF2) covering the left-hand half of the camera and the other one (BF3) covering the right-hand half. A typical image recorded by the camera for the gate time $500 {\mu}$s is shown in Fig. \ref{Fig2}c. In the setup the BF2 filter is centered at $780$ nm (FWHM $10$ nm, transmission $>94\%$) and the BF3 filter is centered at $850$ nm (FWHM $40$ nm, transmission $>95\%$). In this situation, the left-hand side pixels can only detect the `reference' photons within the $10$ nm band around the chosen wavelength, and the right-hand side pixels can detect photons from a broader wavelength range, involving the matching wavelengths. In each experiment, we choose a group of pixels (red squares in Fig.~\ref{Fig2}c and d) as the `reference' detector. To demonstrate SPDC correlations, we have measured the value of the normalized second-order intensity correlation function $g^{(2)}(\vec{r})$, $\vec{r}$ being the position vector, for the chosen reference detector and all the other pixels of the camera. The result, shown in Fig.~\ref{Fig2}d for the threshold $S_{th}=80$, directly exposes the area where the correlated photons could be found. The corresponding values of $g^{(2)}$ are as high as $20$, showing a good ratio between the numbers of real and accidental coincidences. The high values of $g^{(2)}$ around the reference detector are due to the cross-talk, and the spikes in the central area of the camera have very large uncertainties, due to the low illumination level, and should not be trusted.

Because in the absolute calibration method, all photons detected by the reference detector must be accessible to the DUT, the latter is chosen larger than the area showing photon correlations (green square in Fig.~\ref{Fig2}d). Also, the bandwidth of the `DUT' filter is broader than the one of the `reference' one (4 times broader in the case shown in Fig.~\ref{Fig2}c). The bandpass filter in the DUT channel is used for preventing the DUT saturation, which occurs in this channel already at relatively low photon flux and short gate times, because of the large size of the DUT. On the other hand, further reducing the photon flux and the gate time led to the insufficient size of the dataset or required too large number of frames.

It should be noted that the absolute calibration method yields the quantum efficiency of the whole optical channel, including the transmissions of all elements from the light source to the detector~\cite{ware2004single}, which is estimated as $88\%$ in total. All the results given further in this paper have been corrected for this $12\%$ loss in the optical channel.

It is important that for the calibration, we choose collinear non-degenerate phase-matching (Fig.~\ref{Fig2}b). The reason is that in this case, a certain solid angle selected by each of the DUT pixels corresponds to a narrower wavelength range than in the case of a non-collinear phase matching. As a result, the saturation of the DUT is reduced.

\section{Results and discussion}

\begin{figure}[htb]
\centering
\includegraphics[width=12cm]{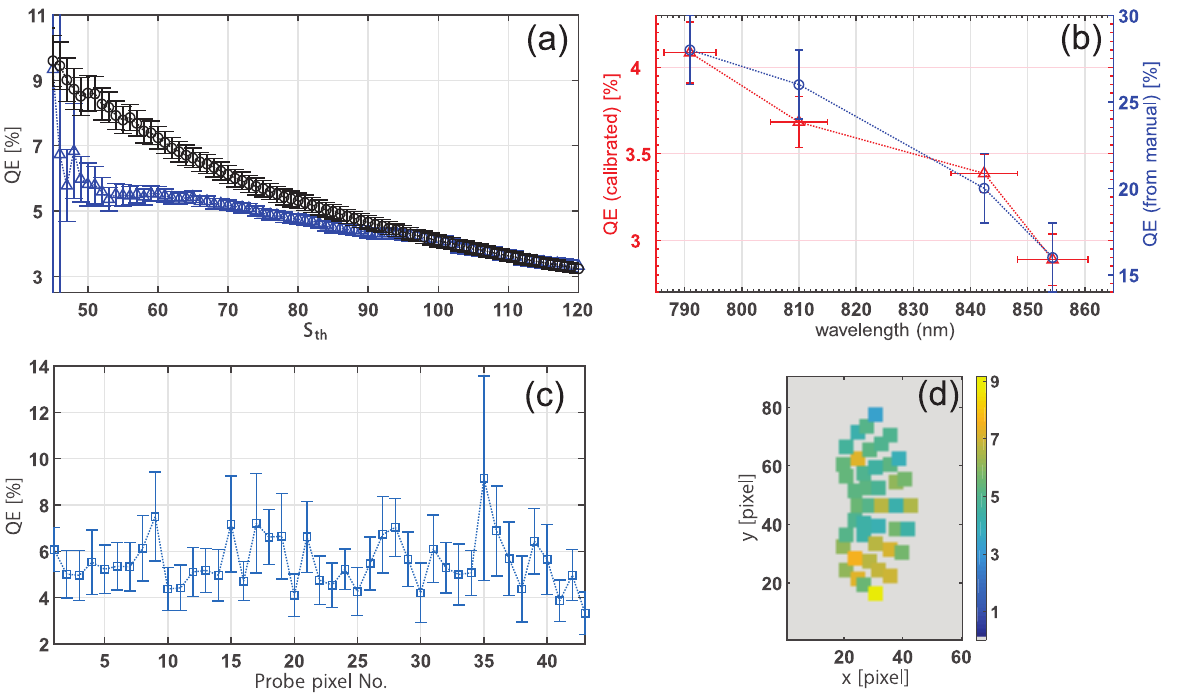}
\caption{(a) The QE measured at the wavelength $790 \textrm{nm} \pm 5 \textrm{nm}$ as a function of the threshold value. Blue triangles are measured through absolute calibration. Black circles are obtained by measuring the mean number of detected photons for a single DUT pixel, with the noise subtracted, as a function of the threshold value, and assuming that the `true' QE value at thresholds above $100$ is correctly measured by the absolute method. (b) Measured QE for the threshold value $100$ as a function of the wavelength (red triangles).  The blue circles show the intensifier QE of the ICCD camera given by the manual. The two datasets show a similar tendency. (c) QE measured for $43$ different pixels for the same threshold $60$ at the wavelength $790 \textrm{nm} \pm 5 \textrm{nm}$. (d) The same data as in panel (c), showing the location of the chosen pixels on the camera matrix.}\label{Fig3}
\end{figure}

Figure~\ref{Fig3}a shows the QE (blue triangles) measured within a range of threshold values from $45$ to $120$ for the wavelength $790~\textrm{nm} \pm 5~\textrm{nm}$, which is conjugated to the wavelength $830$ nm of the chosen reference filter (FWHM $10$ nm, transmission $>95\%$), and filtered by a broadband filter (central wavelength $800$ nm, FWHM $40$ nm, transmission $>99\%$) on the DUT side. As one would expect, the measured QE decreases as the threshold increases. At threshold values smaller than $45$, the measurement error is too large because the incident photons cannot be distinguished against the strong background noise (see Fig.~\ref{Fig1}b).

It is worth noting that, as long as the QE is measured in an absolute way for one threshold value, its measurement for the other threshold values can be performed in a much easier way, as a relative measurement. For this, the mean number of photon count events per single pixel within the DUT, after subtracting the background noise, for the same acquired dataset, is measured as a function of the threshold value (black circles in Fig.~\ref{Fig3}a). The QE resulting from this relative measurement is rescaled in such a way that at high threshold values (we chose a value of $100$), where the SNR reaches a constant value (inset in Fig.~\ref{Fig1}b), it agrees with the result of absolute calibration. Comparing the two curves in Fig.~\ref{Fig3}a, we see that the absolute measurement underestimates the QE at lower threshold values. This could be explained by the DUT saturation, not only by the signal, but also by the background noise. Therefore, at these threshold values it is the result of the relative measurement that should be considered as correct. In experiments where a smaller number of pixels are used as the single-photon detector this problem of saturation is not relevant.

The maximal QE value measured this way is $8.5\% \pm 0.5\%$ at a threshold of $50$. This value is considerably lower than the one given for the intensifier ({\it Unigen II filmless Gen III, Princeton Instruments}) for the same wavelength, namely $28\%$~\cite{PI2015Manual}. This is due to the limited detection probability inside the photocathode-MCP-CCD channel, as well as the fact that any threshold value reduces the QE.

With the same setup, we have measured the dependence of the QE on the wavelength, by changing different reference narrowband filters. In total, we used $4$ filters, each one with $10$ nm FWHM bandwidth, centered at $770$ nm, $780$ nm, $810$ nm, and $830$ nm. In each case, the corresponding DUT wavelengths were covered by one of the two bandpass filters: $800$ nm filter (FWHM $40$ nm, transmission $>99\%$) and $850$ nm filter (FWHM $40$ nm, transmission $>98\%$). The obtained values of QE are plotted in Fig. \ref{Fig3}b (red triangles) for the threshold $S_{th}=100$. The wavelength dependence of the QE is in a good agreement with the intensifier QE dependence (blue circles in Fig. \ref{Fig3}b)~\cite{PI2015Manual}. Clearly, the QE values measured with this relatively high threshold are much lower than the intensifier QE. This is caused by the fact that the overall QE reduces considerably with the threshold value (Fig.\ref{Fig3}a), and is therefore much less than expected from the properties of the intensifier alone.

Finally, Fig. \ref{Fig3}c shows the distribution of the QE over different pixels for the threshold $60$ at the wavelength $790~\textrm{nm} \pm 5~\textrm{nm}$. Figure~\ref{Fig3}d shows the quantum efficiencies measured for $43$ pixels using the absolute calibration method. Within the measurement error, no considerable inhomogeneity of the QE is noticeable.

\section{Conclusion}

In conclusion, using the absolute calibration method, we have measured the quantum efficiency of an ICCD camera. Although such cameras are now often used for single-photon detection in quantum optics experiments, no absolute measurement of quantum efficiency using the Klyshko method has been performed so far. Such a measurement is especially important because a single-photon detection event in an ICCD is distinguished from the noise by setting a threshold, and the value of the threshold has a strong effect on the quantum efficiency. We have performed the absolute calibration of the camera in an autonomous regime, so that different parts of the camera were used as the detector under test and the reference detector. The QE has been measured as a function of the threshold value and the wavelengths. Because at low threshold values, application of the method leads to the saturation of the group of pixels used as the detector under test, we also made a relative measurement of the quantum efficiency. The maximal value observed from the absolute measurement for the wavelength $790~\textrm{nm} \pm 5~\textrm{nm}$ was around $7.0\% \pm 1.5\%$ while the relative measurement resulted in $8.5\% \pm 0.5\%$ in this case. In addition, the inhomogeneity of the camera was assessed by measuring the QE of different pixels. Our results have revealed a large discrepancy between the QE of the photocathode and the total QE of single-photon detection, caused by the thresholding procedure.

\end{document}